\documentclass[10pt,aps,prd,preprintnumbers,twocolumn,nofootinbib]{revtex4-1}
\pdfoutput=1
\usepackage{graphicx,amsfonts,color,comment,amsmath,hyperref,float,times}
\usepackage{capt-of}
\usepackage{amssymb}	

\usepackage{mathrsfs,amssymb}  
\usepackage{cancel}
\usepackage[normalem]{ulem}
\usepackage[dvipsnames]{xcolor}
\usepackage{tikz}
\usepackage[utf8]{inputenc}

\newcommand{\be}{\begin{equation}}
\newcommand{\ee}{\end{equation}}
\newcommand{\bea}{\begin{eqnarray}}
\newcommand{\eea}{\end{eqnarray}}

\newcommand{\dd}{\; \mathrm{d}}
\newcommand{\isotope}[2]{${}^{#2}$#1}

\begin{document}


\title{How blind are underground and surface detectors to strongly interacting Dark Matter?}

\author{Timon Emken}
\email[E-mail: ]{emken@cp3.sdu.dk}
\affiliation{CP3-Origins, University of Southern Denmark, Campusvej 55, DK-5230 Odense, Denmark}

\author{Chris Kouvaris}
\email[E-mail: ]{kouvaris@cp3.sdu.dk}
\affiliation{CP3-Origins, University of Southern Denmark, Campusvej 55, DK-5230 Odense, Denmark}

\date{July 18, 2018}
\begin{abstract}
\noindent 
Above a critical dark matter-nucleus scattering cross section any terrestrial direct detection experiment loses sensitivity to dark matter, since the Earth crust, atmosphere, and potential shielding layers start to block off the dark matter particles. This critical cross section is commonly determined by describing the average energy loss of the dark matter particles analytically. However, this treatment overestimates the stopping power of the Earth crust. Therefore the obtained bounds should be considered as conservative. We perform Monte Carlo simulations to determine the precise value of the critical cross section for various direct detection experiments and compare them to other dark matter constraints in the low mass regime. In this region we find parameter space where typical underground and surface detectors are completely blind to dark matter. This ``hole" in the parameter space can hardly be closed with an increase in the detector exposure. Dedicated surface or high-altitude experiments may be the only way to directly probe this part of the parameter space.
\end{abstract}
\preprint{CP3-Origins-2018-007 DNRF90}
\maketitle

\maketitle

\section{Introduction}
\label{sec:introduction}

The existence of large quantities of dark matter (DM) in the universe is backed by strong astrophysical evidence~\cite{Bertone:2004pz,Ade:2015xua}. Direct detection experiments are consequently trying to observe nongravitational interactions between DM particles from the galactic halo and target atoms inside a terrestrial detector~\cite{Goodman:1984dc,Drukier:1986tm}. So far no conclusive signal has been reported, and detectors continue probing  weaker and weaker DM-nucleon interactions by increasing their exposure~\cite{Aprile:2017iyp} or extend their search to lower DM masses by decreasing their recoil energy threshold~\cite{Petricca:2017zdp,Angloher:2017sxg}. Other approaches focus on experimental signals other than nuclear recoils to probe light DM, as e.g.  DM-electron scatterings~\cite{Essig:2011nj} or inelastic DM-atom scatterings that produce photons~\cite{Kouvaris:2016afs,Ibe:2017yqa}.

These experiments are typically located underground, beneath $\sim$ 1 km of rock, and equipped with additional shielding layers in order to reduce background signals. This comes at the price of impairing the experiment's sensitivity to strongly interacting DM, since elastic DM-nucleus collisions occur not only in the target material but also inside the Earth crust prior to reaching the detector. These collisions slow down and deflect the DM, attenuating the flux of DM particles capable of triggering the detector, such that the Earth crust could shield off DM. Above a critical cross section the shielding effect of the crust or the atmosphere is effective enough that any detector on Earth becomes blind.

The scenario of strongly interacting DM\footnote{We emphasize that strongly interacting DM refers to DM-nucleon interactions and not DM self-interactions.} is highly constrained by astrophysical observations. Strong interactions between DM and baryons in the early universe lead to momentum transfer between them and alter the anisotropy of the Cosmic Microwave background (CMB)~\cite{Chen:2002yh,Dvorkin:2013cea,Gluscevic:2017ywp}. Strong interactions between DM and cosmic ray nuclei lead to the production of neutral pions and consequently $\gamma$-rays~\cite{Cyburt:2002uw,Mack:2012ju,Hooper:2018bfw}. They also introduce a collisional damping effect during cosmological structure formation~\cite{Boehm:2004th}. Further constraints on strongly interacting DM are set by satellite experiments such as IMP7/8~\cite{SnowdenIfft1990,Wandelt:2000ad}, early experiments on the Skylab space station~\cite{Shirk1978,Wandelt:2000ad}, searches for new forces between nucleons~\cite{Fichet:2017bng}, the rocket-based X-ray Quantum Calorimeter experiment (XQC)~\cite{Erickcek:2007jv}, as well as balloon-based experiments such as IMAX~\cite{McGuire:1994pq,Wandelt:2000ad} and the one by Rich et al.~\cite{Rich:1987st}. Considering that terrestrial detectors lose sensitivity above a certain cross section, an important question is whether or not there exists allowed parameter space, a ``window'' between the constraints mentioned above and the ones from direct detection~\cite{Starkman:1990nj}.

Indeed there have been several allowed windows in the constraints on strongly interacting DM over a wide range of DM masses~\cite{Starkman:1990nj,Zaharijas:2004jv}. In order to close these windows, new constraints were derived based on DM capture and heat flow inside the Earth~\cite{Mack:2007xj}, on observations by the IceCube experiment~\cite{Albuquerque:2010bt} and the Fermi Gamma-ray Space Telescope~\cite{Mack:2012ju}. For light DM a window in the constraints was closed by reanalyzing old results from DAMIC and XQC~\cite{Mahdawi:2017cxz}, and using the new results of the CRESST 2017 surface run~\cite{Davis:2017noy}. Most recently several holes in the constraints on super-heavy DM were closed using direct detection experiments only, namely by a reanalysis of CDMS-I~\cite{Kavanagh:2017cru}.

As we mentioned, predetection DM-nuclei scatterings can reduce and effectively eliminate the flux of detectable DM particles in underground detectors. Consequently terrestrial experiments constrain DM-nucleon cross section within a band only. Below the lower limit, there is simply not enough exposure for an experiment to probe DM. Above the upper critical cross section, no constraint can be imposed, because DM interacts too strongly with nuclei and loses enough energy or gets deflected, resulting in a non-detectable flux at the location of the detector. 

In this paper, we use Monte Carlo (MC) techniques to precisely determine this upper critical cross section, above which each experiment fails to probe DM. Previous works on this matter treated the stopping of DM in an overburden with analytic formulas which capture the average energy loss of DM particles traversing through the Earth crust~\cite{Starkman:1990nj,Kouvaris:2014lpa,Foot:2014osa,Davis:2017noy,Kavanagh:2017cru}. However, this analytic treatment does not suffice for a precise determination of the critical cross section. Since the analytic description overestimates the stopping power of the Earth crust by capturing the average energy loss, the bounds obtained this way turn out to be conservative, and may be extended towards higher cross sections. For this purpose dedicated MC simulations of DM particles propagating and scattering in the shielding layers are necessary, as first applied in this context in ~\cite{Zaharijas:2004jv}, to DM-electron scattering experiments in~\cite{Emken:2017erx}, and further developed, motivated and applied in~\cite{Mahdawi:2017cxz,Mahdawi:2017utm}.

Similar simulations have been used to study diurnal signal modulations for intermediate cross sections in~\cite{Collar:1992qc,Collar:1993ss,Hasenbalg:1997hs} and more recently with our \textsc{DaMaSCUS} code in the context of light DM~\cite{Emken:2017qmp,Emken2017a}. Built upon this tool we developed a dedicated simulation code for the determination of the critical cross section of strongly interacting DM. In this paper we present results for CRESST-II~\cite{Angloher:2015ewa}, the CRESST 2017 surface run~\cite{Angloher:2017sxg}, and XENON1T~\cite{Aprile:2017iyp}, and put the direct detection constraints into context with constraints from other sources. We also compare the MC methods to analytic treatments.  

In the first section of this paper we review the commonly applied analytic methods to describe nuclear stopping of DM inside matter, its application to direct detection, and motivate the use of MC simulations. Then we move on to describe our simulations and discuss the method to obtain the critical DM-nucleon cross section. In section~\ref{sec:results} we present our findings, before we conclude in section~\ref{sec:conclusions}. Furthermore we provide the interested reader with a set of appendices, which go into more detail about computational methods, the considered detectors and the implemented model of the Earth crust and atmosphere.

The code \textsc{DaMaSCUS-Crust} developed for this paper is publicly available~\cite{Emken2018a}.

\section{Review of Analytic Methods}
\label{sec:methods}

Different analytic methods have been applied to set limits on the sensitivity of DM detectors on strongly interacting DM. In this section we review these methods, pointing out their shortcomings and motivate the use of MC simulations.

We start by considering a DM particle of mass $m_{\chi}$ and energy $E_\chi$ moving through matter interacting with the nuclei via spin-independent contact interactions. The average energy loss due to elastic scatterings is described~\cite{Starkman:1990nj,Kouvaris:2014lpa} by
\begin{align}
	\frac{\dd E_{\chi}}{\dd x} = -\sum_{i}n_i \int\limits_{0}^{E_R^{\rm max}}\dd E_R\; E_R \frac{\dd \sigma^{\rm SI}_{\chi i}}{\dd E_R}\, , \label{eq: analytic stopping}
\end{align}
where $i$ runs over the abundant nuclei species, $n_i$ is their corresponding number density, and $E_R$ is the nuclear recoil energy with its maximum value $E_R^{\rm max} = \frac{2\mu_{\chi i}^2v_{\chi}^2}{m_i}$. From this point on $\mu_{ij}$ denotes the reduced mass of particle $i$ and $j$, and $m_i$ is the nuclear mass of isotope $i$ with mass number $A_i$. The differential cross section\footnote{To avoid notation clutter we omit the ``SI'' from now on.} is given by
\begin{subequations}
\begin{align}
	\frac{\dd \sigma_{\chi i}}{\dd E_R} &= \frac{\sigma_{\chi i}^{\rm tot}(0)F_i^2(q^2)}{E_R^{\rm max}}\, , 
	\intertext{with}
	\sigma_{\chi i}^{\rm tot}(0) &= \sigma_{\chi n}\frac{\mu_{\chi i}^2}{\mu_{\chi n}^2}A_i^2\, .
\end{align}
\end{subequations}
Here we introduced the spin-independent DM-nucleon scattering cross section~$\sigma_{\chi n}$, and the nuclear form factor $F_i(q^2)$, a function of the momentum transfer $q$, which is related to the recoil energy via $q^2 = 2m_iE_R$. For light DM setting $F_i(q^2)\approx 1$ is an excellent approximation, and the differential cross section no longer depends on $E_R$. Hence in this case eq.~\eqref{eq: analytic stopping} reads
\begin{align}
	\frac{\dd E_{\chi}}{\dd x} = -\sum_i n_i \frac{2\mu_{\chi i}^2\sigma_{\chi i}^{tot}}{m_i m_{\chi}}E_\chi\, ,
\end{align}
and becomes solvable for a homogenous medium with constant density and composition. Then DM particles of initial energy $E^{\rm ini}_{\chi}$ travelling a distance $d$ through this medium will end up with the average final energy
\begin{subequations}
\label{eq:energyloss}
\begin{align}
	 E^{\rm fin}_\chi (d) &= E^{\rm ini}_{\chi} \exp \left[ -\sum_i n_i \frac{2\mu_{\chi i}^2\sigma_{\chi i}^{\rm tot}}{m_i m_{\chi}} d\right]\\
&= E^{\rm ini}_{\chi} \exp \left[- \frac{2\rho \sigma_{\chi n}}{m_{\chi}\mu_{\chi n}^2}\left(\sum_i\frac{f_i \mu_{\chi i}^4A_i^2}{m_i^2}\right)d \right]\, .
\end{align}
\end{subequations}
In the last expression $\rho$ is the (constant) mass density of the medium and $f_i$ the mass fraction of nuclei with mass number $A_i$. In an earlier work we numerically confirmed, that this equation indeed describes the average energy loss accurately~\cite{Emken:2017erx}.

\subsection*{Method a:}
We can use the analytic description of nuclear stopping to give a first estimate of the critical cross section. Every direct detection experiment has a recoil energy threshold $E_R^{\rm thr}$. For a given DM mass $m_{\chi}$ this corresponds to a minimal energy $E_{\chi}^{\rm min}$, such that less energetic DM particle can no longer trigger the detector,
\begin{align}
	E_{\chi}^{\rm min}= \frac{m_Tm_\chi }{4\mu_{\chi T}^2}E_R^{\rm thr}\, . 
\end{align}
Here $m_T$ is the mass of the detector's lightest target nuclei. Above a certain cross section even the most energetic DM particles of the halo travelling towards the detector on the most direct path get slowed down below this value via nuclear stopping. It may serve as a first estimate of the critical cross section. Therefore we solve
\begin{align}
	E_{\chi}^{\rm fin}(d) = E_{\chi}^{\rm min}
\end{align}
for $\sigma_{\chi n}$ using the initial energy $E_{\chi}^{\rm ini} = \frac{m_\chi}{2}\left(v_{\rm esc}+v_{\oplus}\right)^2$ with the maximum DM speed of the halo, the sum of the galactic escape velocity $v_{\rm esc} \approx 544\text{ km/sec}$ and the Earth's velocity in the galactic frame, $v_{\oplus}\approx 230\text{ km/sec}$. The detector under consideration is assumed to be located at an underground depth $d$. We therefore assumed that the particles take the shortest way from the Earth surface to the detector. The resulting estimate of the critical cross section is given by
\begin{align}
	\sigma_{\chi n} = \frac{m_{\chi}\mu_{\chi n}^2}{2\rho d \sum_i\frac{f_i \mu_{\chi i}^4A_i^2}{m_i^2}}\log\left(\frac{E_\chi^{\rm ini}}{E^{\rm min}_\chi}\right)\, . \label{eq: speed criterion}
\end{align}
This method has been developed and applied in \cite{Starkman:1990nj,Kouvaris:2014lpa,Emken:2017erx}. It turned out to be a reasonably good first estimate of the critical cross section for strongly interacting DM, but has a number of shortcomings.
\begin{enumerate}
	\item Unlike the usual lower limit on the cross section set by a direct detection experiment, the critical cross section obtained this way is completely independent of the experimental exposure. In order to set the upper and lower constraint limits on the cross section on equal footing, we need to find the critical cross section using equivalent methods.
	\item To set $v_{\chi}^{\rm ini}=v_{\rm esc}+v_\oplus$ seems rather arbitrary and ignores the knowledge of the halo's speed distribution. However, this makes this approach more conservative.
	\item We assumed $F_i(q^2)\approx 1$ for the nuclear form factor, which holds for light dark matter only.
	\item The actual particles do not travel on a straight line toward the detector. This approach ignores deflections of the particles. However, one can argue that also this makes~\eqref{eq: speed criterion} more conservative since longer distances travelled underground increase the energy loss.
	\item The analytic stopping equation overestimates the stopping power in finding the precise constraint, as pointed out in~\cite{Mahdawi:2017cxz,Mahdawi:2017utm}. We emphasized that the analytic stopping equation accurately describes the average energy loss of DM particles travelling through matter. It does not describe the few, rare particles from the distribution's tails, which scatter on much fewer nuclei than the average particle. The number of these particles is naturally suppressed, which is compensated by the large probability to trigger the detector, once a DM particle actually reaches it. The high cross section both increases the energy loss in the overburden as well as the detection event rates.
\end{enumerate} 
These points make it hard to evaluate precisely the validity of eq.~\eqref{eq: speed criterion}, since point 2 and 4 underestimate the crust's shielding power, while point 5 overestimates it. 

With analytic means problem 5 can hardly be solved, while point 4 can be addressed in the single scattering regime only~\cite{Kavanagh:2016pyr}, where the cross section is too low to impair the overall sensitivity. However, point 1, 2 and 3 can be addressed directly by implementing the analytic stopping description into the computation of detection event rates, as we will do next.

\subsection*{Method b:}
The differential event rate of a generic direct detection experiment is given~\cite{Lewin:1995rx} by
\begin{align}
	\frac{\dd R_i}{\dd E_R} = n_T \frac{\rho_\chi}{m_\chi}\mkern-32mu\int\limits_{v>v_{\rm min}(E_R)}\mkern-32mu\dd v \; v f_d(v) \frac{\dd \sigma_{\chi i}}{\dd E_R}(E_R,v)\, ,\label{eq: dRdER}
\end{align}
where $f_d(v)$ is the speed distribution at the detector's location\footnote{We are not interested in directional detectors at this point and already integrated out the angular part of the full velocity distribution $f(\vec{v})$.}. It is straightforward to include the effect on nuclear stopping by altering the speed distribution. For this purpose we rewrite eq.~\eqref{eq:energyloss} in terms of speed,
\begin{subequations}
\begin{align}
	v^{\rm fin}_\chi (d) &= v^{\rm ini}_{\chi} \exp \left[- \frac{\rho \sigma_{\chi n}}{m_{\chi}\mu_{\chi n}^2}\left(\sum_i\frac{f_i \mu_{\chi i}^4A_i^2}{m_i^2}\right)d \right]\, , \label{eq: deceleration equation}\\
	&\equiv v^{\rm ini}_{\chi} \exp \left[- \Delta d\right]
\end{align}
\end{subequations}
Next we need to find the DM speed distribution function~$f_d(v)$ at depth~$d$ based on our knowledge of the distribution~$f(v)$ at the surface. Since we assume that all particle move on a straight line from the surface to the detector while getting decelerated, the particle flux is conserved,
\begin{align}
	f_d(v_\chi^{\rm fin})v_\chi^{\rm fin}\dd v_\chi^{\rm fin} &= f(v^{\rm ini}_{\chi}) v^{\rm ini}_{\chi} \dd v^{\rm ini}_{\chi}\nonumber\\
 &= \exp \left[2 \Delta d\right]f(\exp \left[ \Delta d\right]v^{\rm fin}_\chi) v^{\rm fin}_\chi \dd v^{\rm fin}_\chi\, .\nonumber\\
	\intertext{We can read off the underground distribution,}
	\Rightarrow f_d(v_\chi^{\rm fin}) &= \exp \left[2 \Delta d\right]f(\exp \left[ \Delta d\right]v^{\rm fin}_\chi)\, .\label{eq:fd}
\end{align}
\begin{figure}[h!]
	\centering
	\includegraphics[width=0.49\textwidth]{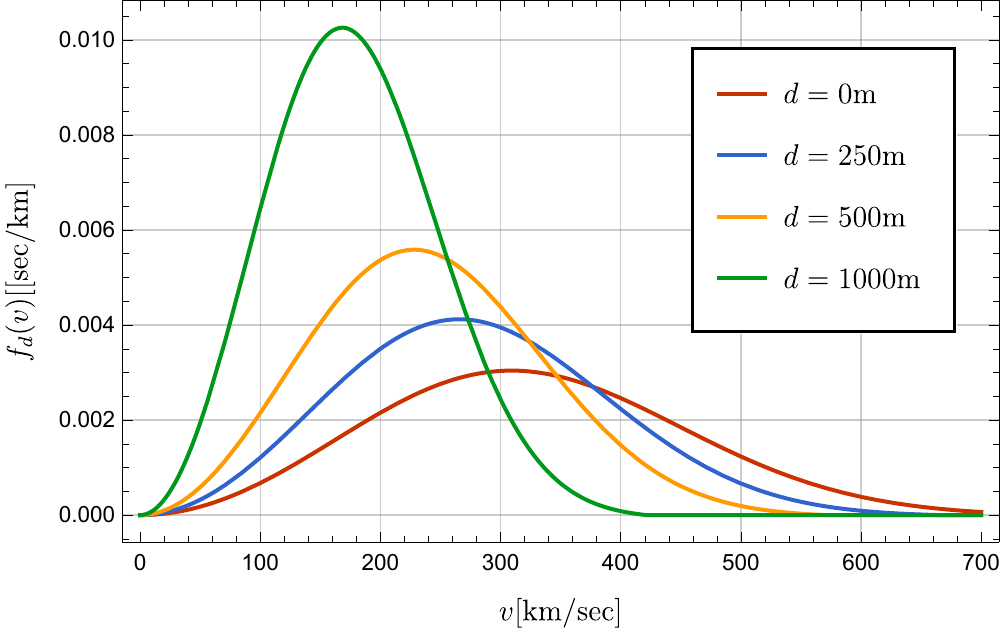}
	\caption{The DM speed distributions given by eq.~\eqref{eq:fd} for $m_\chi=1\text{ GeV}$ and $\sigma_{\chi n}=10^{-30}\text{cm}^2$ at different underground depths $d$. Note the assumption that all particles move directly from the surface straight down to the given depth. Because the particle flux is conserved, the density grows as the particles get slower. This is reflected by the fact that the distribution function is no longer normalized.}
	\label{fig: speeddistribution}
\end{figure}
The speed distribution $f_d(v)$ can be seen in fig.~\ref{fig: speeddistribution}. This expression can be substituted into~\eqref{eq: dRdER}. The resulting spectrum can be written in terms of the surface distribution $f(v)$,
\begin{align}
	&\frac{\dd R_i}{\dd E_R}(d) = n_T \frac{\rho_\chi}{m_\chi}\mkern-32mu\int\limits_{v> e^{\Delta d}v_{\rm min}(E_R)}\mkern-50mu\dd v \; v f(v) \frac{\dd \sigma_{\chi i}}{\dd E_R}(E_R,e^{- \Delta d}v)\label{eq: dRdERd}\, .
\end{align}
The result can then be used to compute recoil spectra and event counts in the usual way, and the shielding effect of the overburden is included automatically, i.e. the signal weakens rapidly with increasing cross section above some critical value. At this point we see that the critical cross section obtained from method~a is exactly the value, above which we obtain zero using eq.~\eqref{eq: dRdERd}. Hence method~a will always slightly overestimate the upper bound of the excluded band compared to method~b.

This method was further refined and applied~\cite{Davis:2017noy}, taking into account the geometry of the whole Earth, not just the overburden above the laboratory as well the atmosphere. However, close to the critical cross section the DM-matter interactions are so strong that virtually all DM particles reach the detector from above. For intermediate cross sections additional attenuation occurs due to DM particles passing through the bulk of the Earth before reaching the detector from below, which is not included in eq.~\eqref{eq:fd}.

Finally the nuclear form factor can also be included, as done in the context of constraining super-heavy DM~\cite{Albuquerque:2003ei,Kavanagh:2017cru}, fixing problem 3 of our previous list. This leaves us with point 4 and 5, both of which can be solved by MC simulations of DM trajectories undergoing scatterings on nuclei of the shielding layers above a detector.

\section{Monte Carlo Simulations}
\label{sec:mc}
\begin{figure*}
	\centering
	\includegraphics{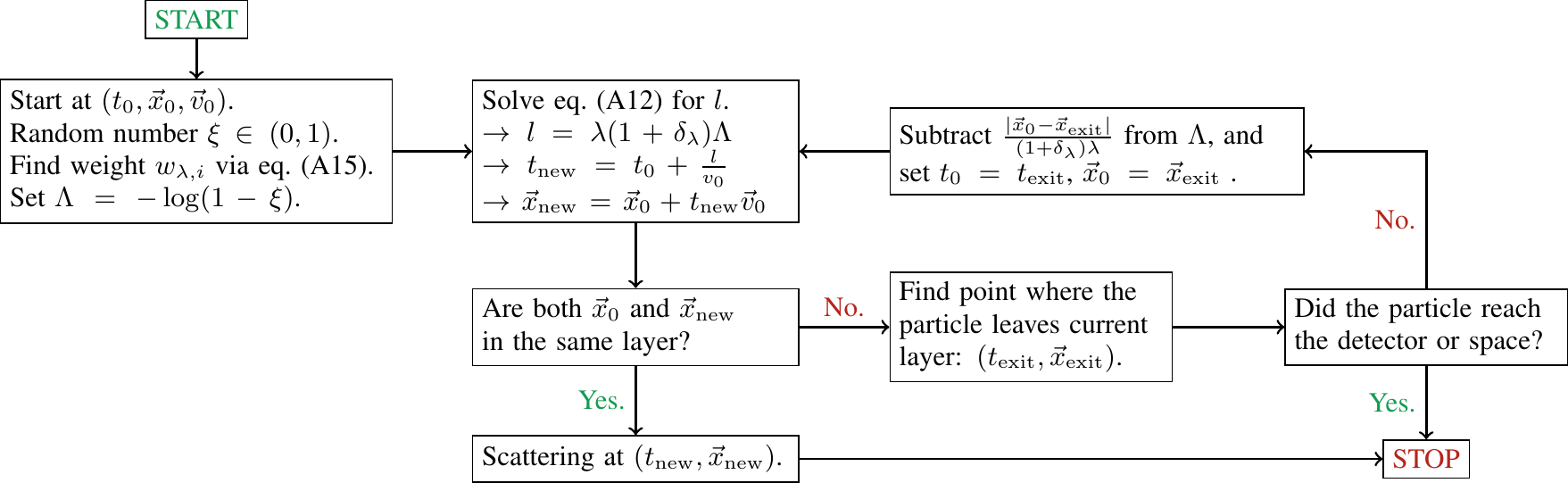}
	\caption{Recursive algorithm to find the location of the next scattering with multiple layers and Importance Sampling (hence the presence of $\delta_\lambda$, see app.~\ref{app:is}). The algorithm concludes when either the location of the next scattering is found, the particle gets reflected into space, or the particle reaches the detector depth. Note that $\lambda$ refers to the local mean free path of a particle at $\vec{x}_0$ with speed $v_0$. It is furthermore important that $\vec{x}_{\rm exit}$ is uniquely identified as element of the new layer.}
	\label{fig:algorithm}
\end{figure*}
The general principles of our MC simulations are laid out in detail in~\cite{Emken:2017qmp}. We simulate DM particles that travel underground on straight lines until they scatter elastically on a nucleus, where the particle gets decelerated and deflected. As opposed to the \textsc{DaMaSCUS} code we model the overburden, i.e. the Earth crust, the atmosphere and shielding layers as planar layers instead of a spherical Earth model. Close to the critical  cross section the DM-baryon interaction is so strong that particles reach the detector virtually exclusively from above. For the details of the crust and atmosphere models we refer to appendix~\ref{app:crust}. 

The number of shielding layers in our simulations is variable, each layer is characterized by its density and nuclear composition. The mean free path~$\lambda$ inside layer $i$ is given by
\begin{align}
\lambda_i^{-1}&= \sum_k \lambda_{ik}^{-1} \equiv \sum_{k}n_{ik}\sigma^{\rm tot}_{\chi k}=\sum_{i}f_{ik}\frac{\rho_i}{m_{k}}\sigma^{\rm tot}_{\chi k}\, . \label{eq:lambdainverse}
\end{align}
Here $\rho_i$ denotes the $i$th layer's mass density, and $n_{ik}$, $f_{ik}$ the corresponding number density and elemental abundance of the isotope $k$. Before a DM particle scatters inside a shielding layer, it travels freely for a distance
\begin{align}
-\lambda_i\log(1-\xi)\;,\quad\text{with }\xi\in[0,1]\, , \label{eq:L}
\end{align}
with $\xi$ being a uniformly distributed random number. Where a particle passes a boundary into another shielding layer before the next scattering, the mean free path changes along the way. To account for this discontinuity, we implement a recursive algorithm to find the location of the next scattering regardless of how many layer boundaries are crossed. A flow chart can be found in figure~\ref{fig:algorithm}.

 Once the location of scattering is determined, the collision on a nucleus is simulated, where the probability of colliding on nucleus species $k$ of layer $i$ is given by
\begin{align}
	P_k = \frac{\lambda_{ik}^{-1}}{\lambda_i^{-1}}\, .
\end{align}
The particle's velocity changes from $\vec{v}_\chi$ to
\begin{align}
\vec{v}_\chi^\prime = \frac{ m_k \left|\vec{v}_\chi\right| \vec{n}+m_\chi \vec{v}_\chi}{m_k m_\chi}\, .\label{eq:vnew}
\end{align}
The unit vector $\vec{n}$ yields the final direction of the DM particle in the center-of-mass frame. Here we implicitly assumed that the target nuclei are particles at rest. The relative velocity is dominated by the incoming DM particle. We denote the angle between $\vec{n}$ and the incoming velocity $\vec{v}_\chi$ as $\alpha$. For light DM the distribution of $\vec{n}$ is isotropic. For more details on the distribution function of $\alpha$, or rather $\cos\alpha$, we refer to the appendix~\ref{app:is}.

The main goal of the MC simulations is to obtain a precise estimate for the DM speed distribution $f(v)$ at the detector's location over the interval of interest $[v_{\rm min},(v_{\rm esc}+v_\oplus)]$. The minimum speed depends on the experiment's threshold~$E_{\rm thr}$, target atoms, and the energy resolution~$\sigma_E$,
\begin{align}
	v_{\rm min} = \sqrt{\frac{m_T (E_{\rm thr}-3\sigma_E)}{2\mu_{\chi T}^2}}\, ,\label{eq:vmin}
\end{align}
where $m_T$ is again the mass of the lightest target isotope. For light DM the interval of interest typically lies in the high velocity tail. 

The above steps repeat until either the particle gets reflected into space, gets decelerated below the minimum speed, or reaches the depth of the detector. In the last case we record its speed.

To make the connection from the data to direct detection rates, we need to estimate the speed distribution $f_d(v)$ at depth~$d$. Our estimate is given by the product of an attenuation factor and the normalized speed distribution,
\begin{align}
	\hat{f}_d(v) =  a\times \hat{f}^{\rm KDE}_d(v)\, .\label{eq:MCf}
\end{align}
The attenuation factor accounts for the fraction of particles making it to detector depth, while still being detectable,
\begin{align}
	a =\frac{\sum\limits_{i=1}^{N} w_i}{N_{\rm tot}}\, .
\end{align}
Note that we only sample initial conditions within the interval of interest $[v_{\rm min},(v_{\rm esc}+v_\oplus)]$ to speed up computations. Hence, if we simulate $N_{\rm sim}$ particles, this corresponds to a total particle number of
\begin{align}
	N_{\rm tot} = \frac{N_{\rm sim}}{\int\limits_{v_{\rm min}}^{v_{\rm esc}+v_\oplus}\dd v\; f(v)}\, .
\end{align}
Here we assumed to have simulated $N_{\rm sim}$ trajectories to get $N$ data points, associated with each is a statistical weight $w_i$. Finally $f^{\rm KDE}_d(v)$ is the normalized Kernel Density Estimate (KDE) of the speed PDF on the interval $[v_{\rm min},v_{\rm esc}+v_\oplus]$, which we describe in detail in app.~\ref{app:kde}.

For a given point in parameter space $(m_\chi,\sigma_{\chi n})$ and a given experiment at depth $d$, the simulations provide us with a sample of speed data and their corresponding weights $\{(v_1,w_1),...,(v_N,w_N)\}$ as well as the attenuation factor $a$. We follow the following steps to extract the recoil spectrum.
\begin{enumerate}
	\item Estimate the speed distribution via eq.~\eqref{eq:MCf} using KDE, see app.~\ref{app:kde}.
	\item Compute $\frac{\dd R}{\dd E_R}$ via numerical integration using eq.~\eqref{eq: dRdER}.
	\item Depending on the experiment of interest compute the recoil spectrum $\frac{\dd R}{\dd E}$ including the detector's energy resolution, efficiency, and quenching factors.
	\item Compute detection signal counts and likelihoods.
\end{enumerate}
For a given DM mass $m_\chi$ we repeat the simulations for different cross sections. Starting at the usual upper bound we systematically increase the cross section until the likelihood grows above (1-CL), and the experiment no longer constrains the cross section. The exact value of the critical cross section is finally determined by interpolating the likelihood as a function of the cross section.

\subsection{Computational details}
The simulation code \textsc{DaMaSCUS-Crust} is written in C++ loosely based on our previous \textsc{DaMaSCUS} tool~\cite{Emken2017a}. It is publicly available~\cite{Emken2018a}. All simulations for this work were run on the Abacus 2.0, a 14.016 core supercomputer of the DeIC National HPC Center, SDU.
At various places we use interpolation with Steffen splines, which has the great advantage of avoiding spurious extrema~\cite{Steffen1990}. All numerical integrations in the simulation and data analysis are performed with the adaptive Simpson method~\cite{Kuncir1962}. For the non-parametric, data-based estimate of the speed distribution function we use univariate Kernel Density Estimation (KDE). To speed up the simulations we implement Importance Sampling (IS) as proposed by Mahdawi and Farrar~\cite{Mahdawi:2017cxz,Mahdawi:2017utm}. The details on both KDE and IS are presented in appendix~\ref{app:computationalmethods}.

\section{Results}
\label{sec:results}

\begin{figure}[h!]
	\centering
	\includegraphics[width=0.49\textwidth]{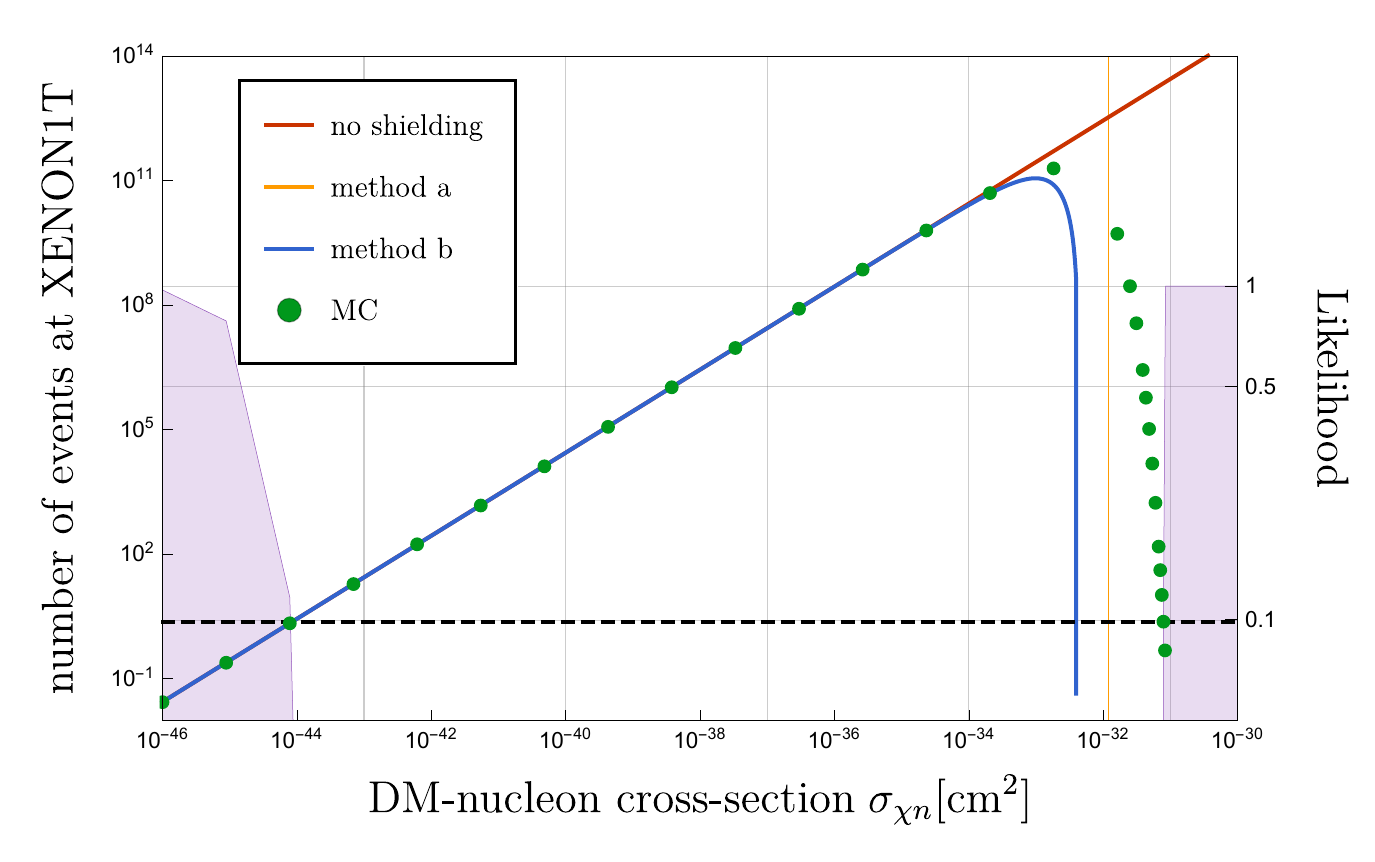}
	\caption{The expected number of events and likelihood at XENON1T as a function of the DM-nucleon scattering cross section for a DM mass of 10 GeV. Note that for intermediate cross sections, particles arriving at the detector from below will also contribute and attenuate event counts, giving rise to diurnal modulations. This is not taken into account here, where all particles are assumed to reach the detector from above and the focus lies on finding the critical cross section of strongly interacting DM.}
	\label{fig:N}
\end{figure}

We compare the different methods reviewed in section~\ref{sec:methods} with results from the MC simulations. As an illustrative example we show the number of events at XENON1T for a DM mass of $m_\chi=10$ GeV as a function of the DM-nucleon scattering cross section in figure~\ref{fig:N}. Without the stopping effect of the 1400m of rock overburden taken into account the number of events naturally keeps growing with the cross section. If, instead of the unmodified Maxwell-Boltzmann distribution, we use eq.~\eqref{eq:fd}, we successfully reproduce the characteristic behaviour. While the overburden's shielding effect has virtually no effect for intermediate cross sections, since the mean free path is much longer than the underground depth, above a critical cross section the event number in the detector drops drastically. By looking for the cross section for which the likelihood drops below 0.1, we obtain the 90\%CL bound.

It should be mentioned at this point that the assumption that all particles reach the detector from above is of course only valid for high cross sections. In the intermediate regime particles entering the detector from below will travel through the bulk of the Earth and undergo scatterings. This leads to additional attenuations and diurnal modulations, which will not show up in fig.~\ref{fig:N}. For the investigation of the intermediate regime, we refer to~\cite{Emken:2017qmp}.

Comparing the blue curve to the MC results, the advantage of the simulation approach becomes very clear. The analytic stopping equation obviously overestimates the stopping power of an overburden and makes the event number drop too fast with increasing cross section. In reality, particles which scatter fewer times than the average still reach the detector capable of triggering it. Therefore MC simulations make constraints on strongly interacting DM more stringent, extending to higher cross sections. The resulting limits are not just more restrictive, but also more accurate, robust and consistent, since upper and lower bounds are on equal footing.

In a recent paper~\cite{Mahdawi:2017utm} the authors claim that the analytic description fails in deriving the critical cross section of strongly interacting DM, quoting a discrepancy in the number of events of multiple orders of magnitude. However, looking at fig.~\ref{fig:N} it is clear that any method which conservatively underestimates the critical cross section, will lead to much higher event numbers compared to the corresponding MC simulations. Yet, this discrepancy says very little about the accuracy of the critical cross section estimate as the actual quantity of interest, since the event number drops very steeply. The limits obtained with the analytic descriptions may be conservative and improvable, but they are still valid. They typically underestimate the critical cross section just by a factor of a few.

For completeness we also include the corresponding bound obtained with method~a, i.e. the simple speed cutoff criterion. In this case it gives a reasonable and conservative estimate, which is more restrictive than the limit of method~b as expected. However, without the MC results a quality assessment would not have been possible, as discussed in sec.~\ref{sec:methods}. 

\begin{figure}
	\centering
	\includegraphics[width=0.49\textwidth]{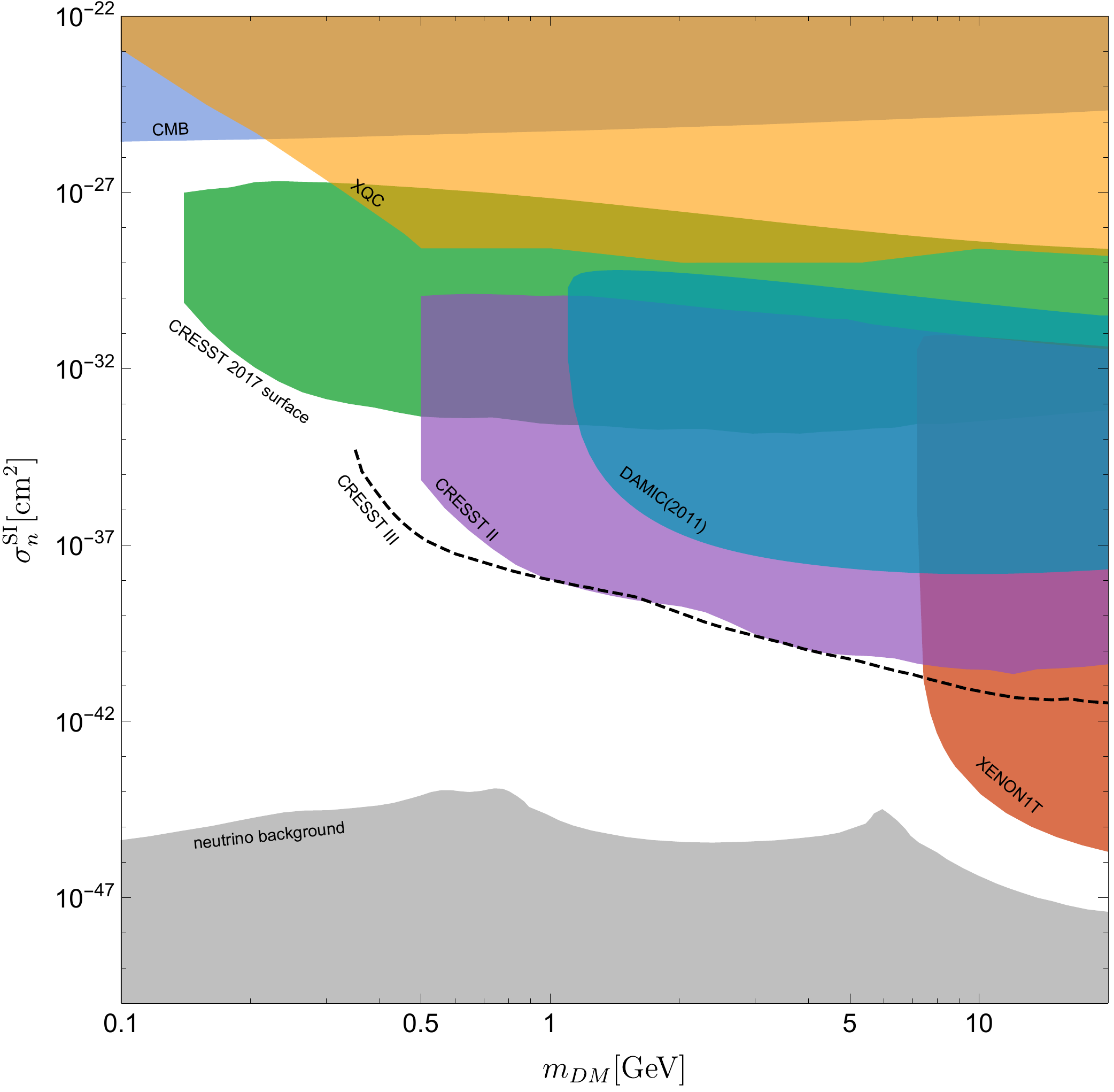}
	\caption{Our results for the 90\% CL constraints on light DM for CRESST-II~\cite{Angloher:2015ewa}, the CRESST 2017 surface run~\cite{Angloher:2017sxg}, DAMIC(2011)~\cite{Barreto:2011zu}, and XENON1T~\cite{Aprile:2017iyp}. Also included are constraints from XQC~\cite{Erickcek:2007jv}, and the CMB~\cite{Gluscevic:2017ywp}. At the bottom of the plot we included the neutrino background~\cite{OHare:2016pjy}, and in black dashed lines we indicate the new constraints from CRESST-III~\cite{Petricca:2017zdp}.}
	\label{fig:constraints}
\end{figure}

We show the main results of this study in fig.~\ref{fig:constraints}, the constraints on DM with masses between 100 MeV and 20 GeV from CRESST-II, XENON1T, DAMIC(2011), and the CRESST 2017 surface run, together with constraints from the XQC experiment and the CMB. For each mass and detector\footnote{For details on the considered detectors we refer to app.~\ref{app:experiments}.}, we obtain an excluded band of cross sections, from a lower limit to the upper critical cross section due to shielding of strongly interacting DM.

The DAMIC(2011) constraints are fully covered by the two experiments of the CRESST collaboration. The purpose of including these result is to compare them to limits obtained with the \textsc{DMATIS} code~\cite{Mahdawi:2017cxz} as an independent and valuable cross check of our simulation. For the masses between 1 and 100 GeV we find an average relative deviation between the two limits of about 15\% with slightly higher deviations for masses of order $\mathcal{O}(1\text{ GeV})$. But overall the two limits seem to agree to a reasonable precision. Further cross checks and comparisons might be desirable, though the \textsc{DMATIS} code has not been released at the time of submission of this paper.

Both CRESST-II and XENON1T are located deep underground at LNGS. Hence it comes to no surprise that they turn out to be rather insensitive to strongly interacting DM. In the low-mass regime they constrain cross sections up to $\sim10^{-30}\text{cm}^2$ and $\sim10^{-31}\text{cm}^2$ respectively.

\begin{figure}[h!]
	\centering
	\includegraphics[width=0.49\textwidth]{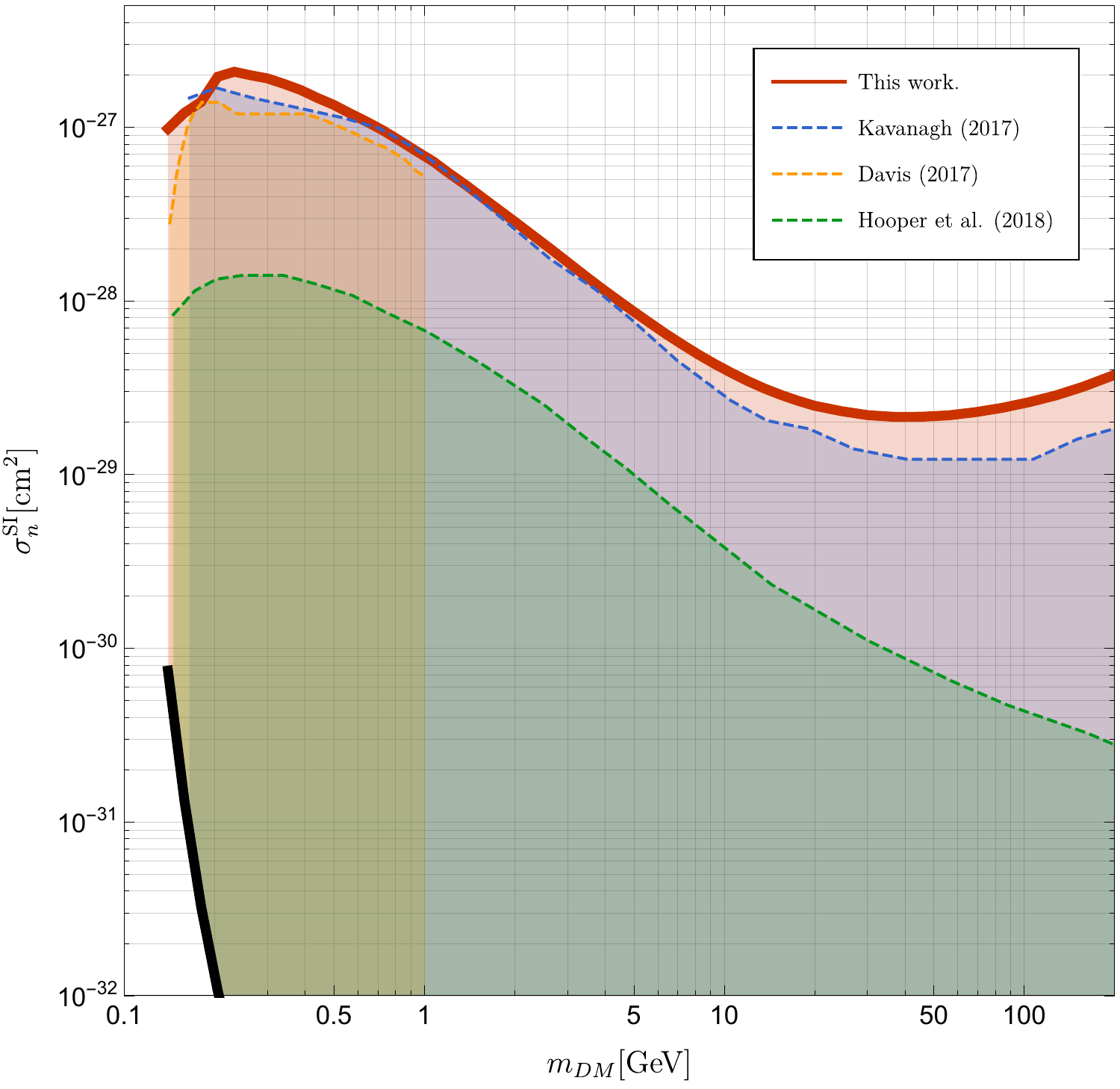}
	\caption{Comparison of the CRESST 2017 surface run constraints on strongly interacting DM between  analytically obtained constraints by Davis (2017)~\cite{Davis:2017noy}, Kavanagh (2017)~\cite{Kavanagh:2017cru}, and Hooper et al. (2018)~\cite{Hooper:2018bfw}, and MC results of this paper.}
	\label{fig:comparison}
\end{figure}

Most interesting is last year's CRESST 2017 surface run of a prototype detector developed for the $\nu$-cleus experiment. As opposed to the vast majority of DM detectors it was not placed underground and is therefore ideal to constrain strongly interacting DM. It probes and constraints cross section around three orders of magnitude higher than XENON1T and CRESST-II despite its small exposure. The resulting constraint close all allowed windows between other terrestrial detectors and the XQC experiments. However, a window between the new constraints and the CMB limits remains, which might get narrowed by taking constraints from cosmic rays into account~\cite{Cyburt:2002uw}.  

In fig.~\ref{fig:comparison} we compare our CRESST 2017 surface run MC constraints with analytic results from recent works. Depending on the DM mass the constraints reported in~\cite{Davis:2017noy,Kavanagh:2017cru}, whose analytic methods are based on method b of this paper, underestimate the critical cross section by up to a factor of 2. It should be mentioned again that this method assumes that a DM particle simply moves along a straight path through the Earth while continuously losing energy, an approximation that breaks down for light DM, as discussed in section~\ref{sec:methods}. Most recently, the authors of ~\cite{Hooper:2018bfw} presented a new, and simple analytic method to establish very conservative limits, which circumvents this assumption. It is based on rescaling the number of events while filtering out DM particles that scatter at least once in the overburden. The constraints are therefore considering only the unscattered DM population to set a constraint. But a single scattering is not sufficient to render the flux of the remaining, scattered DM particles undetectable, which is why their constraints are so conservative. The constraints from MC simulations improve upon these results by at least one order of magnitude.

We should emphasize that the exposure has only little effect on the critical cross section as opposed to the lower limit. Despite an exposure larger by a factor of $\sim$700, the bounds for masses below 20 GeV by XENON1T exceed the ones by CRESST-II by $\sim$10\% only. Comparing to DAMIC(2011) and the CRESST surface run(2017), it becomes clear that the underground depth is by far the dominating factor. This means that in order to close the window between the CMB and CRESST 2017 surface run constraints by means of direct detection, a dedicated surface or high-altitude run of a sensitive portable detector is needed.

\section{Conclusions}
\label{sec:conclusions}
In this paper we determine for the first time the precise limits that different underground detectors set on the DM cross section-mass parameter space focusing in the interesting and to great extent unconstrained low DM mass region. We determine the precise critical DM-nucleon scattering cross section with MC techniques above which a given detector with a certain depth becomes blind to DM. This is important because it is essential to know if there are blind spots in the DM parameter space not covered by terrestrial detectors. In such a case surface or high-altitude runs of detectors may close remaining windows in the parameter space. 

 We presented a brief review of the analytic methods, described the details of our MC simulations and presented updated limits for CRESST-II, XENON1T, and the CRESST surface run(2017). In addition we compared the MC results with the analytic methods confirming that the critical cross section estimated via the analytic methods presented in~\cite{Kouvaris:2014lpa} is typically within a factor of a few below the precise MC one.
 
We found that the recent surface run of the prototype detector by the CRESST collaboration constrains strongly interacting DM with cross sections up to about $2 \times 10^{-27}\text{cm}^2$.  This closes the allowed window for light DM between the constraints by terrestrial detectors and the XQC experiment, as also shown in~\cite{Davis:2017noy}. However there is still an open window between the CMB constraints and the ones of the CRESST 2017 surface run. It is open for masses between 140 and about 300 MeV and extends over more than an order of magnitude in cross section. If DM lies within this window, we find that all underground detectors are completely blind to DM almost regardless of how much they can improve in exposure. Dedicated surface or preferably high-altitude experiments could close this window. Furthermore, a number of complementary astrophysical constraints to direct detection exist as listed in the introduction, which could exclude the detectors' blind spots. Constraints from cosmic rays narrow the allowed window~\cite{Cyburt:2002uw}. Under the assumption that DM can annihilate into Standard Model particles, the constraints from the anomalous heat flow in the Earth rule out the upper left corner of figure~\ref{fig:constraints} and close the small window in parameter space~\cite{Mack:2007xj}. Naturally, these constraints do not apply to e.g. asymmetric DM. The \textsc{DaMaSCUS-Crust} code we developed for this purpose can be used for any future experiment of a given target, depth, exposure and energy threshold, providing a definite and precise answer on what parameter space can be excluded.

Finally, we should also emphasize that DM with mass beneath the reach of CRESST could also have large enough cross sections, such that underground scatterings have to be taken into account. Hence, future efforts should be put towards performing a similar analysis for DM-electron scattering experiments. For models with a dark-photon portal a large hierarchy between the DM-electron and the DM-nucleon scattering cross section follows, such that nuclear stopping becomes very relevant for detectors probing DM-electron interactions. Furthermore, it would be desirable in this context to weaken the assumption of a heavy mediator, since scenarios with light mediators are much less constrained. In conclusion we hope that this paper and the corresponding code will serve as a tool to specify the extend to which direct detection experiments constrain models with strong DM-nucleus interactions.

\begin{acknowledgments}
The CP$^3$-Origins centre is partially funded by the Danish National Research Foundation, grant number DNRF90, and by the Danish Council for Independent Research, grant number DFF 4181-00055. Computation/simulation for the work described in this paper was supported by the DeIC National HPC Centre, SDU.
\end{acknowledgments}

\appendix

\section{Computational Methods}
\label{app:computationalmethods}
\subsection{Univariate Kernel Density Estimation}
\label{app:kde}
A central problem for the data treatment in this study is the question how to estimate an unknown probability density function (PDF) $f(x)$ with domain $I$, i.e. $x\in I$, based on a data set $\{x_1,...,x_N\}$ and possibly weights $\{w_1,...,w_N\}$ in a non-parametric way. Histograms are the first straight-forward idea that comes to mind, but there is a more sophisticated method, which has the advantage of producing a continuous and smooth estimate of the true PDF, Kernel density estimation (KDE)~\cite{Rosenblatt1956,Parzen1962}. The PDF can be estimated via
\begin{align}
	\hat{f}_h(x) = \frac{1}{h \sum_i w_i}\sum_{i=1}^{N} w_i K\left(\frac{x-x_i}{h}\right)\, . \label{eq: kde}
\end{align}
The parameter $h$ is called the bandwidth. An applicable kernel $K(x)$ satisfies
\begin{subequations}
	\begin{align}
	&K(x)\geq 0\, , \forall x\in I\, ,\\
	&\int_I \dd x \; K(x) = 1\, ,\\
	&\int_I \dd x \;x K(x) = 0\, .
	&\end{align}
\end{subequations}
Commonly used kernels include uniform, triangular, Epanechnikov (parabolic), cosine, and many more. We usually choose a simple Gaussian kernel,
\begin{align}
	K(x) = \frac{1}{\sqrt{2\pi}}\exp\left(\frac{-x^2}{2}\right)\, . \label{eq:gauss kernel}
\end{align}
We also define the scaled kernel,
\begin{align}
	K_h(x) &\equiv \frac{1}{h}K\left(\frac{x}{h}\right)\, ,
	\intertext{ such that}
	\hat{f}_h(x) &= \frac{1}{\sum_i w_i}\sum_{i=1}^{N} w_i K_h\left(x-x_i\right)\, .\label{eq: kde2}
\end{align}
The bandwidth $h$ is the only free parameter and its choice is therefore crucial, similarly to the bin width for histograms. Choose $h$ too large and the KDE $\hat{f}_h(x)$ might oversmooth crucial features of the true PDF $f(x)$. Choose it too small and we might resolve statistical fluctuations. One estimate is Silverman's rule of thumb~\cite{silverman1986density},
\begin{align}
	h = \left(\frac{4}{3 N}\right)^{1/5}\hat\sigma\, ,
\end{align}
which takes the sample's size $N$ and standard deviation $\hat\sigma$ into account.

\begin{figure}[h!]
	\centering
	\includegraphics[width=0.48\textwidth]{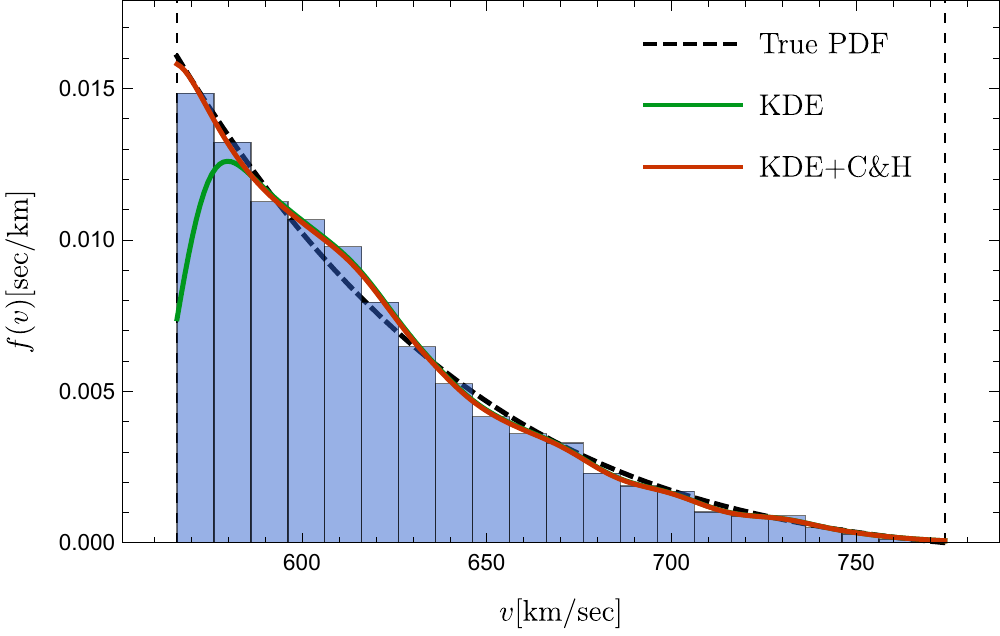}
	\caption{Kernel density estimation of a small data set (N=5000) on a bounded domain. Clearly the standard KDE underestimates the PDF close to the lower boundary. Once the pseudodata method of Cowling and Hall is implemented, the bias vanishes.}
	\label{fig: cowling and hall}
\end{figure}

The Gaussian Kernel~\eqref{eq:gauss kernel} might be problematic since the kernel domain is $(-\infty,\infty)$. It is a well-known and extensively studied problem of KDE that this can introduce a significant error in cases, where the domain $I$ is bounded. The estimate given by~\eqref{eq: kde2} underestimates the true PDF close to the boundary. This is due to the fact, that the Kernel does not contain information of the boundary and assigns weight to the region beyond the boundary, where naturally no data is found, as visible in the green line of fig.~\ref{fig: cowling and hall}.

The issue is most severe for large probability mass close to the boundary, i.e. if the true PDF does not vanish at edges. This is typically the case for the PDF we estimate in the MC simulations of this work. Many methods of boundary bias removal have been proposed, see e.g.~\cite{Karunamuni2005} and references therein. An effective and easy to implement example is to simply reflect the data around the boundary. Say we have a PDF $f(x)$ with support $\left[x_{\rm min},x_{\rm max}\right]$ and $f(x_{\rm min})> 0$, and a data sample $\left\{x_1,...,x_N\right\}$ with $x_{\rm min}\leq x_i \leq x_{\rm max}$. We could now adjust eq.~\eqref{eq: kde2} via
\begin{align}
	\hat{f}_h(x) = \frac{1}{\sum_i w_i}\sum_{i=1}^{N} w_i \left[K_h\left(x-x_i\right)+K_h\left(x-x_i^{\rm refl}\right)\right]\, ,\label{eq: kde with reflection}
\end{align}
with $x_i^{\rm refl}\equiv 2x_{\rm min}-x_i$. This illustrates how the generation of pseudo-data beyond the boundary can reduce boundary bias. This specific example however has the drawback that it leads to $\hat{f}^\prime(x_{\rm min})=0$, which might not be a feature of the true PDF. Cowling and Hall proposed a procedure to generate pseudodata without this disadvantage~\cite{Cowling1996}. There are ways to linearly combine data points from within the support into pseudodata points $(x_{(-1)},...,x_{(-m)})$ beyond the boundary, in a way that the correct edge behaviour of $f(x)$ is reproduced. The corrected KDE is then
\begin{subequations}
\begin{align}
	\hat{f}_h(x) &=\frac{1}{\sum_i w_i}\times\nonumber\\
	&\hspace{-0.85cm}\left[\sum_{i=1}^{N} w_i K_h\left(x-x_i\right) + \sum_{i=1}^m w_{(-i)}K_h(x-x_{(-i)})\right]\, . \label{eq: kde with pseudo data}
\intertext{One way of combining data into pseudo data is the following three-point-rule,}
	x_{(-i)} &= 4 x_{\rm min} - 6 x_i + 4 x_{2i} -x_{3i}\, ,\\
	w_{(-i)} &= \frac{w_i+w_{2i}+w_{3i}}{3}\, ,\\
	 m&=\frac{N}{3}\, .
\end{align}
\end{subequations}
Since the normalization of~\eqref{eq: kde with pseudo data} on its support is no longer guaranteed, it might be necessary to renormalize the PDF. The result can be seen in fig.~\ref{fig: cowling and hall}.

\subsection{Importance sampling}
\label{app:is}
A challenge to MC simulations arises, if the desired sample is composed of rare events. In our scenario it might be the case, that in order to get a single DM speed data point at detector depth we would have to simulate millions and millions of particles. This is, in the best case, inconvenient and inefficient, since it requires a tremendous amount of computing power. In the worst case MC simulations might not be applicable at all. This happens typically if the regions of interest of the involved PDFs fall into their tails. Introducing Importance Sampling~(IS), a standard technique in rare event MC simulations, can soften this problem considerably\footnote{For an introduction to IS in rare event simulations, we refer to~\cite{Bucklew:2004}.}. The basic idea is to artificially increase the number of rare events in a controlled manner, compensating by a proper weighting factor.

Assuming for simplicity that the simulations involve a single probability density function $f(x)$, which determines the behavior of every single particle trajectory. However, our final data set consists exclusively of particles that make it to the detector depth while still being detectable. Hence, we aggressively filtered out trajectories or events based on a given criterion. These events can be very rare, and their statistical properties are expected to differ from the `typical' particle, most of which had been filtered out. Consequently, the underlying PDF $g(x)$ of the data set differs from $f(x)$. 

A MC simulation with IS makes use of this, in order to increase the chance of these rare events. Instead of $f(x)$, we change the simulation PDF to some new PDF $\hat{g}(x)$, which should approximate $g(x)$, mimicking the PDF of the successful events during the simulations, and therefore rare events occur more frequently. The introduced bias has to be compensated by a data weight function, as we describe below. If $\hat{g}(x)$ is chosen poorly, the method becomes unstable. It is crucial that the biased PDFs approximate the distributions of the final data set. This is why we ran consistency checks and compared ``brute force'' simulations with IS simulations for several examples.

The weighting factor can be understood based on a simple observation. Assume we have a random variable $X$ with PDF $f_X(x)$ and we are interested in the expectation value of a quantity $Y(X)$ on a given interval $I$,
\begin{subequations}
	\begin{align}
	\langle Y\rangle_I &= \int_I\dd x \; Y(x) f(x)\, .
\intertext{We can trivially rewrite this as}
&= \int_I \dd x\; Y(x) \frac{f(x)}{\hat{g}(x)}\hat{g}(x)\, .
\end{align}
\end{subequations}
The function $\hat{g}(x)$ can be regarded as the new PDF and the factor $\frac{f(x)}{\hat{g}(x)}$ as the weighting function. Consequently, we sample from $\hat{g}(x)$ instead of $f(x)$ during the simulations. Provided that we chose $\hat{g}(x)$ as described above, the desired rare events will be favoured, and we can significantly reduce the amount of MC runs necessary to gather the data sample we need. 

In many cases this method makes MC simulations practically feasible, where normal MC simulations would fail due to lack of time and resources. This makes the method particularly useful to simulate strongly interacting DM particles moving through the Earth crust and shielding layers~\cite{Mahdawi:2017cxz,Mahdawi:2017utm}. Detectable DM particles which make it to the detector depth turn out to differ in their statistical properties from the average DM particle in two ways:
\begin{enumerate}
	\item They scatter fewer times, and travel freely further than the mean free path.
	\item They scatter more in the forward direction.
\end{enumerate}
Both points  make the survival to the detector depth more likely. This also indicates how we should alter the PDFs of our MC simulations. In the following, we describe these modifications to the PDFs and the corresponding weight functions in detail.

The distance $x$ a DM particle travels through matter without scattering on one of its constituents has the underlying PDF
\begin{align}
	f_{\lambda}(x) = \frac{1}{\lambda}\exp\left(-\frac{x}{\lambda}\right)\, ,
\end{align}
with the mean free path $\lambda$. By increasing the mean free path, we introduce a bias that favors DM trajectories that can make it to the detector with enough energy to get detected. We therefore chose the altered PDF
\begin{align}
	g_\lambda(x)= \frac{1}{(1+\delta_\lambda)\lambda}\exp\left(-\frac{x}{(1+\delta_\lambda)\lambda}\right)\, ,
\end{align}
with $\delta_\lambda>0$. Every time we sample a distance $l_i$ from this PDF by solving
\begin{align}
	&\int\limits_{0}^{l_i}\dd x\; g_\lambda(x) = \xi\, ,\label{eq:MCsampling}
\end{align}
where $\xi\in(0,1)$ is a uniform random number, we keep track of the weighting factor
\begin{align}
	w_{\lambda,i} = \frac{f_\lambda(l_i)}{g_\lambda(l_i)} = (1+\delta_\lambda)\exp\left(-\frac{\delta_\lambda l_i}{(1+\delta_\lambda)\lambda}\right)\, .
\end{align}
If the trajectory consists of $n_S$ scattering events, the overall weight of the trajectory is
\begin{align}
	w_\lambda = \prod_{i=0}^{n_S} w_{\lambda,i}\, .
\end{align}
Especially when the particle freely passes different shielding layers a more useful expression for the weight is
\begin{align}
	w_{\lambda,i} = (1+\delta_\lambda)(1-\xi)^{\delta_{\lambda}}\, .\label{eq:weight}
\end{align}

Next we turn to the scattering angle, where we want to favour forward scattering. For light DM we can approximate the nuclear form factor $F_A(q^2)\approx 1$ and the scattering is isotropic in the center of mass frame. In other words $\cos \alpha$ is a uniform random quantity with PDF
\begin{align}
	f_{\alpha}(\cos\alpha) = \frac{1}{2}\, ,
\end{align}
and domain $[-1,1]$. Since forward scattering favors detectable particles at detector depth, we introduce the following IS biased PDF
\begin{align}
	g_\alpha(\cos\alpha) = \frac{1+\delta_\alpha\cos\alpha}{2}\, ,\quad\text{with }\delta_\alpha\in[0,1]\, .\label{eq: is pdf ldm}
\end{align}
At each scattering we sample from $g_\alpha$, i.e. we obtain $\cos \alpha_i$ by generating a uniform random number $\xi\in(0,1)$ and solving
\begin{align}
	&\int\limits_{-1}^{\cos\alpha_i}\dd\cos\alpha\; g_\alpha(\cos\alpha) = \xi\, ,\\
	\Longleftrightarrow &\cos\alpha_i = \frac{-1 + \sqrt{(1-\delta_\alpha)^2+4\delta_\alpha\xi}}{\delta_\alpha}\, .
\end{align}
Note that $\lim\limits_{\delta_\alpha\rightarrow 0}\cos\alpha_i = 2\xi - 1$ as expected. We keep track of the weighting factor,
\begin{align}
	w_{\alpha,i} &= \frac{f_\alpha(\cos\alpha_i)}{g_\alpha(\cos\alpha_i)} = \frac{1}{1+\delta_\alpha \cos\alpha_i}\nonumber\\
	& = \left[ (1-\delta_\alpha)^2 +4\delta_{\alpha}\xi\right]^{-1/2}\, .
\end{align}
For heavier DM the approximation for the nuclear form factor $F_A(q^2)\approx 1$ no longer holds. Then the scattering is not isotropic in the center-of-mass frame anymore. Instead the PDF for the scattering angle $\cos\alpha$ of a DM particle of speed $v_\chi$ scattering on a nucleus with mass number $A$ is given by
\begin{align}
	f_{\alpha}(\cos\alpha;\, v_\chi)=\frac{F^2_{A}(q^2)}{\int\limits_{-1}^{+1}\dd(\cos\alpha^\prime)\;F_A^2(q^{\prime 2})}\, ,
\end{align}
where $q^2=2\mu_{\chi A}^2v_\chi^2(1-\cos\alpha)$.
We implemented the Helm form factor~\cite{Helm:1956zz} in our code, which is given by
\begin{subequations}
\begin{align}
	F_A(q)&=3\left(\frac{\sin (q r_n)}{(q r_n)^3}-\frac{\cos(q r_n)}{(q r_n)^2} \right)\exp\left(-\frac{q^2 s^2}{2}\right)\, ,\\
	\intertext{with }r_n &=\sqrt{c^2+\frac{7}{3}\pi^2a^2-5s^2}\, ,\\
	c&=\left(1.23 A^{1/3}-0.6\right)\text{fm}\, ,\\
	a&=0.52\text{ fm}\, ,\\
	s&=0.9\text{ fm}\, .
\end{align}
\end{subequations}
In analogy to eq.~\eqref{eq: is pdf ldm}, the IS biased PDF is chosen to be
\begin{align}
	g_{\alpha}(\cos\alpha;\, v_\chi)=f_{\alpha}(\cos\alpha;\, v_\chi)+\frac{\delta_\alpha}{2}\cos\alpha\, ,
\end{align}
in order to favour forward scattering. In this case a slight subtlety enters, since the inclusion of the loss of coherence already favors forward scattering naturally. Therefore we have to be careful that $g_\alpha(\cos\alpha;v_\chi)$ does not yield negative values for backwards scattering. This occurs if
\begin{align}
	\delta_\alpha > 2 f_{\alpha}(-1;\, v_\chi)\, .
\end{align} 
If this is the case for a particular scattering we adjust $\delta_\alpha$ to zero for this scattering, and set the corresponding weight $w_{\alpha,i}$ to 1 correspondingly.

The overall weight factor of a successful trajectory with $n_S$ scatterings is then
\begin{align}
	w_\alpha = \prod_{i=1}^{n_S}w_{\alpha,i}\, .
\end{align}
If we use IS for both involved PDFs, for $l$ and $\cos\alpha$, the overall weight of a data point is $w = w_{\lambda}w_\alpha$.

\section{Description of direct detection experiments}
\label{app:experiments}
In this appendix we review the details of the different detectors relevant for the analysis of this work.
\subsection{CRESST-II}
The phase 2 of the CRESST-II experiment set leading limits on light DM for masses down to 0.5 GeV~\cite{Angloher:2015ewa,Angloher:2017zkf}. The target material of the Lise module consists of 300g of $\text{CaWO}_4$ crystals. CRESST-II was located at the LNGS under 1400m of rock. It realized an energy threshold of $E_{\rm thr} = 307\text{ eV}$ and an energy resolution of $\sigma_E=62\text{ eV}$. The theoretical recoil spectrum is given by
\begin{align}
\frac{\dd R}{\dd E} &= \int \dd E_R\; \text{Gauss}(E|E_R,\sigma_E)\sum_{i}\epsilon_i(E)\frac{\dd R_i}{\dd E_R}	\, ,\label{eq: dRdE}
\end{align}
where we substitute eq.~\eqref{eq: dRdER}. In the acceptance region of $[E_{\rm thr},40\text{ keV}]$ 1949 events survived all cuts, and the energy data was released in~\cite{Angloher:2017zkf}, together with the cut survival probability $\epsilon_i(E)$ for each of the three targets. The exposure corresponded to 52.15 kg days. Using Yellin's maximum gap method~\cite{Yellin:2002xd} to compute the likelihood of a given point in parameter space reproduces the 90\% CL constraints on light DM accurately.

Meanwhile the results from the phase 1 of CRESST-III have been published~\cite{Petricca:2017zdp}, and provide new, leading constraints on light DM. Compared to CRESST-II the threshold was lowered to $\sim 100\text{ eV}$, such that the experiment probes DM as light as 350 MeV. With an exposure of 2.39 kg days 33 events were detected in the acceptance region. Since the recoil data has not been released by the CRESST collaboration at the time this manuscript was prepared, we only include full results for CRESST-II. However, due to their similarity the inclusion of new result will be straightforward as soon as the new data gets released. 
\subsection{CRESST 2017 surface run}
In July 2017 the CRESST collaboration presented results from a surface run of a prototype detector developed for the $\nu$-cleus experiment with a 0.49g sapphire ($\text{Al}_2\text{O}_3$) target~\cite{Angloher:2017sxg}. Despite a small net exposure of only 0.046 g days this experiment is perfect for this work, since it was set up in a surface building of the Max Planck Insitute in Munich with only $\sim$ 30cm of concrete ceiling and the Earth atmosphere as shielding. With a remarkable low threshold of $E_{\rm thr}\approx 19.7$ eV and energy resolution $\sigma_E\approx 3.74$ eV it constraints DM mass as low as 140 MeV. The number of observed events in the region of interest $[E_{\rm thr},600\text{ eV}]$ was 511, and the cut efficiency was conservatively set to 1. The energy data of the 511 events is made public as ancillary files along with the preprint on arXiv. Otherwise the analysis is the same as used for CRESST-II, using eq.~\eqref{eq: dRdE} and Yellin's maximum gap method, which reproduces the official constraints up to a factor of $\sim 2$.

\subsection{XENON1T}
The XENON collaboration published their first results from XENON1T in May 2017~\cite{Aprile:2017iyp}. The results obtained with an exposure of 35.6 ton days were consistent with the background-only hypothesis. The nuclear recoil energy threshold was $E_{\rm thr} = 5\text{ keV}$, such that XENON1T probes DM heavier than a few GeV. Its region of interest extends from the threshold to 40 keV. Just as the CRESST-II experiment, the XENON1T detector is set up at LNGS underneath 1400m of rock.

We implemented a simplified analysis for XENON1T based on eq.~\eqref{eq: dRdER} with an assumed flat efficiency of 82\% across the region of interest.
\subsection{DAMIC(2011)}
The DAMIC experiment uses silicon CCDs to search for light DM~\cite{Barreto:2011zu}. The results from an engineering run using a 0.5g target in 2011 yield stronger constraints on strongly interacting DM than later DAMIC runs, because it was operated at a relatively shallow underground site at Fermilab with a depth of only around 350 feet($\sim$106.7m), and lead shielding of 6 inches ($\sim$15cm). Even though the excluded area due to DAMIC(2011) is already covered by CRESST-II and the CRESST surface run (2017) we include it in our analysis. Comparing this result to the ones by Mahdawi and Farrar~\cite{Mahdawi:2017cxz}, who use equivalent simulations for specifically this experiment, is a valuable cross-check and validation of our and their results. 

The DAMIC(2011) run had an exposure of 0.107 kg days and a threshold of $E_{\rm thr}=0.04\text{ keV}_{\rm ee}$ and observed 106 events in a region of interest of $[E_{\rm thr},2\text{ keV}_{\rm ee}]$. To match the results of~\cite{Mahdawi:2017cxz} we compute the likelihood and 90\% CL constraints via Poisson statistics. The region of interest in terms of the nuclear recoil energy is chosen to be~$[0.55\text{ keV},7\text{ keV}]$.
 
\begin{table*}
\begin{minipage}[b]{0.45\textwidth}
\centering
\begin{tabular}{|l|l|c|l|l|}
			\cline{1-2}\cline{4-5}
			Element	&[wt\%]	&& Element	&[wt\%]\\
			\cline{1-2}\cline{4-5}
			\isotope{O}{16}	&	46.6	&&	\isotope{N}{14}	&	75.6\\
			\cline{1-2}\cline{4-5}
			\isotope{Si}{28}	&	27.7	&&	\isotope{O}{16}	&	23.1\\
			\cline{1-2}\cline{4-5}
			\isotope{Al}{27}	&	8.1	&&	\isotope{Ar}{40}	&	1.3\\
			\cline{1-2}\cline{4-5}
			\isotope{Fe}{56}	&	5.0	&&	\isotope{C}{12}	&	0.02\\
			\cline{1-2}\cline{4-5}
			\isotope{Ca}{40}	&	3.6	&&	\isotope{Ne}{20}	&	0.001\\
			\cline{1-2}\cline{4-5}
			\isotope{Ka}{39}	&	2.8	&&		&	\\
			\cline{1-2}
			\isotope{Na}{23}	&	2.6	&&		&	\\
			\cline{1-2}
			\isotope{Mg}{24}	&	2.1	&&		&	\\
			\cline{1-2}\cline{4-5}
			\textbf{Total}	&	98.5	&&	\textbf{Total}	&	100\\
			\cline{1-2}\cline{4-5}
		\end{tabular}
		\caption{Nuclear composition of the Earth crust (left) and atmosphere (right)}
		\label{tab:composition}	
	\end{minipage}\hfill
\begin{minipage}[b]{0.5\textwidth}
\centering
		\includegraphics[width=\textwidth]{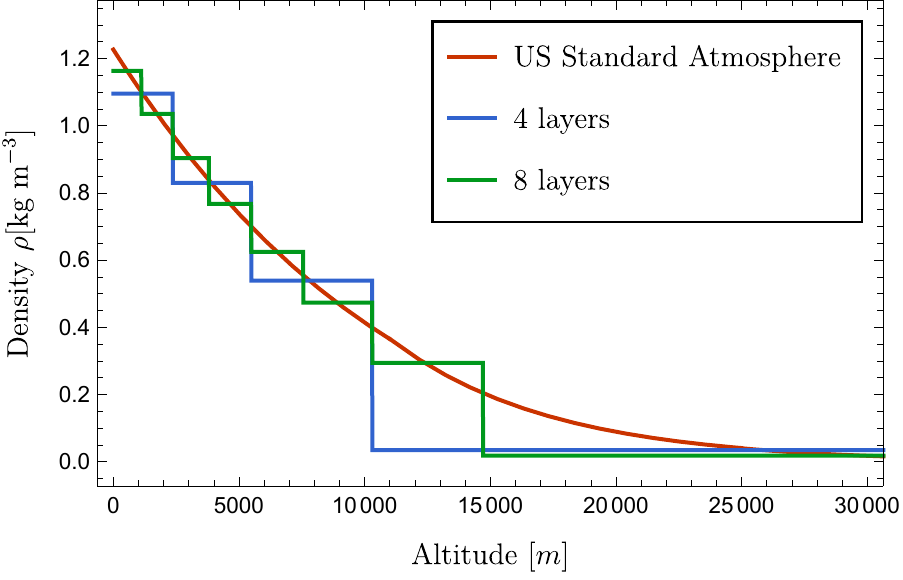}
		\captionof{figure}{Density profile of the US Standard Atmosphere model}
		\label{fig:usatmosphere}	
\end{minipage}
\end{table*}
\section{Modeling the Earth crust and atmosphere}
\label{app:crust}

We model the Earth crust as a layer of constant mass density $\rho=2.7\text{ g cm}^{-3}$, consisting of nuclei, whose abundances are given on the left hand side of table~\ref{tab:composition}~\cite{Rudnick20031}. 

In order to account for the atmosphere we implemented the US Standard Atmosphere (1976)~\cite{atmosphere}, which extends to an altitude of 86km. The composition is listed on the right hand side of table~\ref{tab:composition}. The density profile is plotted in fig.~\ref{fig:usatmosphere}. For the sake of easy implementation we divide the atmosphere in a set of layers with constant density, such that the integral $\int\rho(x)\dd x$ is the same for each layer.  As a cross-check we varied the number of layers, to ensure that our results are stable and do not depend on this discretization.

\bibliographystyle{JHEP}
\bibliography{library.bib} 

\end{document}